\documentclass[12pt]{article}
 \usepackage{epsfig}
  \usepackage{wrapfig}
   \usepackage{graphicx}
\begin{document}

\begin{titlepage}
\begin{center}

{\Large\bf Problems with the perturbative QCD \\[2mm]
interpretation of HERMES data on semi-inclusive \\[2mm]
lepto-production of pions}

\end{center}
\vskip 2cm
\begin{center}
{\bf Elliot Leader}\\
{\it Imperial College London\\ Prince Consort Road, London SW7
2BW, England }
\vskip 0.5cm
{\bf Alexander V. Sidorov}\\
{\it Bogoliubov Theoretical Laboratory\\
Joint Institute for Nuclear Research, 141980 Dubna, Russia }
\vskip 0.5cm
{\bf Dimiter B. Stamenov \\
{\it Institute for Nuclear Research and Nuclear Energy\\
Bulgarian Academy of Sciences\\
Blvd. Tsarigradsko Chaussee 72, Sofia 1784, Bulgaria }}
\end{center}

\vskip 0.3cm
\begin{abstract}
\hskip -5mm

A theoretical analysis has been performed of the HERMES
semi-inclusive deep inelastic data on the difference between
$\pi^{+}$ and $\pi^{-}$ multiplicities. It turns out that the
application of a standard perturbative QCD analysis, using, as
usual, isospin symmetry for the fragmentation functions, leads to
a poor description of the deuteron data at small $z$ values. If
one allows a breaking of isospin invariance, a good fit to both
proton and deuteron data can be achieved for all measured $z$, but
the level of isospin violation, especially at small values of $z$,
is simply not credible. We suspect that the problem is a
consequence of using the factorized QCD treatment in a kinematic
region where it is unjustified.


\end{abstract}

\end{titlepage}

\newpage
\setcounter{page}{1}

\section{Introduction}

Semi-inclusive deep inelastic production of hadrons in
lepton-nucleon collisions plays a key role in the determination of
polarized sea quark densities. In the perturbative QCD approach,
valid in the kinematic region of large momentum transfer to the
nucleon, the cross-section to produce a hadron $h$ is expressed,
in leading order  (LO), as a product of parton densities $q(x)$,
functions of Bjorken-$x$,  and fragmentation functions $D_q^h(z)$,
reflecting the probability of a quark $q$ to fragment into the
hadron $h$ carrying a fraction $z$ of its energy. This factorized
form is modified into a convolution when going beyond LO and, of
course, the parton densities and fragmentation functions (FFs),
develop a slow logarithmic dependence on the scale $Q^2=-q^2$,
where $q^\mu$ is the 4-momentum transfer from the lepton to the
nucleon.

In studying the latest HERMES data on $\pi^\pm$ production on
protons and deuterons \cite {HERMES} we have found that the
application of the standard perturbative QCD analysis to the
differences of the multiplicities
$M_{p(d)}^{\pi^{+}-\pi^{-}}(x,Q^2,z)$ using, as usual, isospin
invariance for the FFs, leads to a poor description of the data at
low $z$. Relaxing the demand for isospin invariance leads to
violations of isospin conservation at a level which is simply not
credible. The violation is biggest at small values of $z$ and this
has led us to suspect that the apparent violations are an artifact
of using the factorized QCD treatment in a kinematic region where
it is unjustified. Indeed, many years ago, Berger \cite{Berger}
proposed criterium for delineating the kinematic regions where the
standard approach should be valid, and we have found that for the
HERMES data, the data bin corresponding to the smallest values of
$z$ i.e. $0.2\leq z \leq 0.3$ lies outside Berger's safe region,
as will be explained in detail in the paper.

\section{QCD treatment of hadron multiplicities}

The multiplicities $M_{p(d)}^{\pi}(x,Q^2,z)$ of pions using a
proton (deuteron) target are defined as the number of pions
produced, normalized to the number of DIS events, and can be
expressed in terms of the semi-inclusive cross section
$\sigma_{p(d)}^{\pi}$ and the inclusive cross section
$\sigma_{p(d)}^{DIS}$:
\begin{eqnarray}
M_{p(d)}^{\pi}(x,Q^2,z)&=&
\frac{d^3N_{p(d)}^{\pi}(x,Q^2,z)/dxdQ^2dz}{d^2N_{p(d)}
^{DIS}(x,Q^2)/dxdQ^2}=\frac{d^3\sigma_{p(d)}^{\pi}(x,Q^2,z)
/dxdQ^2dz}{d^2\sigma_{p(d)}^{DIS}(x,Q^2)/dxdQ^2}\nonumber\\
&=&\frac{(1+(1-y)^2)2xF_{1p(d)}^{\pi}(x,Q^2,z)+2(1-y)xF_{Lp(d)}
^{\pi}(x,Q^2,z)}
{(1+(1-y)^2)2xF_{1p(d)}(x,Q^2)+2(1-y)F_{Lp(d)}(x,Q^2)}.
\label{M_exp_th}
\end{eqnarray}
In Eq. (\ref{M_exp_th}) $F_1^{\pi}, F_L^{\pi}$ and $F_1, F_L$ are
the semi-inclusive and the usual nucleon structure functions,
respectively. $F_1^{\pi}$ and $ F_L^{\pi}$ are expressed in terms
of the unpolarized parton densities and  fragmentation functions,
while $F_1$ and $F_L$ are given by the unpolarized parton
densities.

It turns out that the pion multiplicities are measured with enough
precision so that one can directly extract from their differences
$M_{p(d)}^{\pi^{+}-\pi^{-}}(x,Q^2,z)$ the nonsinglet FFs
$D_q^{\pi^{+}-\pi^{-}}(z,Q^2) =
(D_q^{\pi^{+}}-D_q^{\pi^{-}})(z,Q^2)$. Knowledge of these is
important to test whether the additional assumptions for the
favored and unfavored FFs which are usually made in the QCD
analyses of the multiplicities $M_{p(d)}^{\pi^{+}}(x,Q^2,z)$ and
$M_{p(d)}^{\pi^{-}}(x,Q^2,z)$, are or are not correct.

Using the charge conjugation invariance of the strong interactions
in the case of the pion fragmentation functions
\begin{equation}
D_q^{\pi^{+}-\pi^{-}}=-D_{\bar q}^{\pi^{+}-\pi^{-}},
~~~D_g^{\pi^{+}-\pi^{-}}=0
\label{C_inv}
\end{equation}
and the assumption $s(x,Q^2)=\bar s(x,Q^2)$ for the strange
unpolarized parton densities, the following expressions for the
proton and deuteron semi-inclusive structure functions hold in NLO
QCD \cite{ChrLeader}\footnote{In this paper formulas are presented
for the semi-inclusive cross section. Note that a factor 1/2 is
missing in the formulas for the deuteron target.}:
\begin{eqnarray}
2F_{1p}^{(\pi^{+}-\pi^{-})}(x,Q^2,z)&=&\frac{1}{9}(4u_v\otimes D_u^
{\pi^{+}-\pi^{-}}+d_v\otimes D_d^{\pi^{+}-\pi^{-}})\otimes(1+\frac
{\alpha_s(Q^2)}{2\pi}C^1_{qq}),\\
F_{Lp}^{(\pi^{+}-\pi^{-})}(x,Q^2,z)&=&\frac{\alpha_s(Q^2)}{2\pi}\frac{1}{9}
(4u_v\otimes D_u^{\pi^{+}-\pi^{-}}+d_v\otimes D_d^{\pi^{+}-\pi^{-}})
\otimes C^L_{qq},\\
2F_{1d}^{(\pi^{+}-\pi^{-})}(x,Q^2,z)&=&\frac{1}{18}(u_v+d_v)
\otimes (4D_u^{\pi^{+}-\pi^{-}}+D_d^{\pi^{+}-\pi^{-}})\otimes (1+
\frac{\alpha_s(Q^2)}{2\pi}C^1_{qq}),\\
F_{Ld}^{(\pi^{+}-\pi^{-})}(x,Q^2,z)&=&\frac{\alpha_s(Q^2)}{2\pi}
\frac{1}{18}(u_v+d_v)\otimes (4D_u^{\pi^{+}-\pi^{-}}+D_d^{\pi^{+}-\pi^{-}}),
\otimes C^L_{qq},
\label{nonsinglF1_FL}
\end{eqnarray}
where $u_v(x,Q^2)$ and $d_v(x,Q^2)$ are the valence unpolarized
parton densities, $C^1_{qq}(x,z,Q^2)$ and $C^L_{qq}(x,z,Q^2)$ are
the Wilson coefficient functions \cite{Wilson_coeff}. Note also
that the arguments $(x,Q^2)$ for the parton densities and
$(z,Q^2)$ for the FFs in the equations above are omitted. The
remarkable properties of these semi-inclusive structure functions
are that the gluon fragmentation function does not contribute into
the structure functions themselves, nor to the $Q^2$ evolution of
the nonsinglet quark fragmentation functions
$D_q^{\pi^{+}-\pi^{-}}(z,Q^2)$.

According to isospin $SU(2)$ symmetry
\begin{equation}
D_d^{\pi^{+}-\pi^{-}}(z,Q^2)= -D_u^{\pi^{+}-\pi^{-}}(z,Q^2)
\label{SU2}
\end{equation}
and then Eqs. (3-6) take the following simple form:
\begin{eqnarray}
2F_{1p}^{(\pi^{+}-\pi^{-})}(x,Q^2,z)&=&\frac{1}{9}(4u_v-d_v)
\otimes(1+\frac{\alpha_s(Q^2)}{2\pi}C^1_{qq})\otimes D_u^{\pi^{+}-\pi^{-}},
\label{F1p}\\
F_{Lp}^{(\pi^{+}-\pi^{-})}(x,Q^2,z)&=&\frac{\alpha_s(Q^2)}{2\pi}\frac{1}{9}
(4u_v-d_v)\otimes C^L_{qq}\otimes D_u^{\pi^{+}-\pi^{-}},
\label{FLp}\\
2F_{1d}^{(\pi^{+}-\pi^{-})}(x,Q^2,z)&=&\frac{1}{6}(u_v+d_v)
\otimes (1+\frac{\alpha_s(Q^2)}{2\pi}C^1_{qq})\otimes D_u^{\pi^{+}-\pi^{-}},
\label{F1d}\\
F_{Ld}^{(\pi^{+}-\pi^{-})}(x,Q^2,z)&=&\frac{\alpha_s(Q^2)}{2\pi}
\frac{1}{6}(u_v+d_v)\otimes C^L_{qq}\otimes D_u^{\pi^{+}-\pi^{-}},
\label{FLd}
\end{eqnarray}
where the unpolarized valence quark densities $q_v(x,Q^2)$ and the
nonsinglet fragmentation function $D_u^{\pi^{+}-\pi^{-}}(z,Q^2)$
satisfy the NLO QCD evolution equations.

In LO QCD approximation the longitudinal proton and deuteron
structure functions $F_{Lp(d)}^{(\pi^{+}-\pi^{-})},~F_{Lp(d)}$ are
equal to zero and we obtain for the differences of the pion
multiplicities:
\begin{eqnarray}
M_{p}^{\pi^{+}-\pi^{-}}(x,Q^2,z)&=&\frac{(4u_v-d_v)(x,Q^2)D^
{\pi^{+}}_{u_{v}}(z,Q^2)}{[4(u+\bar u)+d+\bar d +2s](x,Q^2) },
\label{diff_MpLO}\\
M_{d}^{\pi^{+}-\pi^{-}}(x,Q^2,z)&=&\frac{3(u_v+d_v)(x,Q^2)D^
{\pi^{+}}_{u_{v}}(z,Q^2)}{[5(u+\bar u+d+\bar d) +4s](x,Q^2) },
\label{diff_MdLO}
\end{eqnarray}
where the notation $D^{\pi^{+}}_{u_{v}}$ is used for the
nonsinglet fragmentation function
\begin{equation}
D^{\pi^{+}}_{u_v} \equiv D_u^{\pi^{+}}-D_{\bar u}^{\pi^{+}}=
D_u^{\pi^{+}-\pi^{-}}.
\label{Duv}
\end{equation}

So, if isospin symmetry holds we have two independently measured
quantities, the multiplicities $M_{p}^{\pi^{+}-\pi^{-}}$ and
$M_{d}^{\pi^{+}-\pi^{-}}$ for the one nonsinglet fragmentation
function $D^{\pi^{+}}_{u_v}(z, Q^2)$ which can thus be determined
in LO QCD directly and independently from the data at measured
values of $Q^2$ without using an input parametrization for
$D^{\pi^{+}}_{u_v}(z, Q^2)$ and its $Q^2$ evolution.

\section{Results}

In our analysis we have used the $[Q^2,z]$ presentation of the
HERMES proton and deuteron data on pion multiplicities
\cite{HERMES}, corrected for exclusive vector meson production.
The pion multiplicities are given for 4 z-bins
[0.2-0.3;~0.3-0.4;~0.4-0.6;~0.6-0.8] as functions of $Q^2$. Note
that to any measured value of $Q^2$ the corresponding value of the
Bjorken variable $x$ is attached. The total number of the
$\pi^{+}$ and $\pi^{-}$ data points for the proton and deuteron
targets is 144, 72 for $\pi+$ and 72 for $\pi-$ data. So, there
are 36 data points (9 for every z-bin) for the multiplicity
$M_{p}^{\pi^{+}-\pi^{-}}(x,Q^2,z)$, as well as for
$M_{d}^{\pi^{+}-\pi^{-}}(x,Q^2,z)$.

\subsection{Isospin SU(2) symmetry}

Our results on the fragmentation function $D^{\pi^{+}}_{u_v}(z,
Q^2)$ extracted from the proton data using Eq. (\ref{diff_MpLO})
and deuteron data using Eq. (\ref{diff_MdLO}) are presented in
Fig. 1, blue and red points, respectively. For the LO parton
densities the CTEQ6l parametrization \cite{CTEQ6l} has been used.
In the calculations of the errors presented in Fig. 1 only the
statistical errors of the multiplicities have been taken into
account. The extracted nonsinglet FFs should coincide within the
errors. As seen from Fig. 1, they are not in agreement for the
first z-bin, and partially for the second one. The use of a
different set of PDFs practically does not change the situation.
\begin{figure} [ht]
\begin{center}
  \includegraphics[height=.46\textheight]{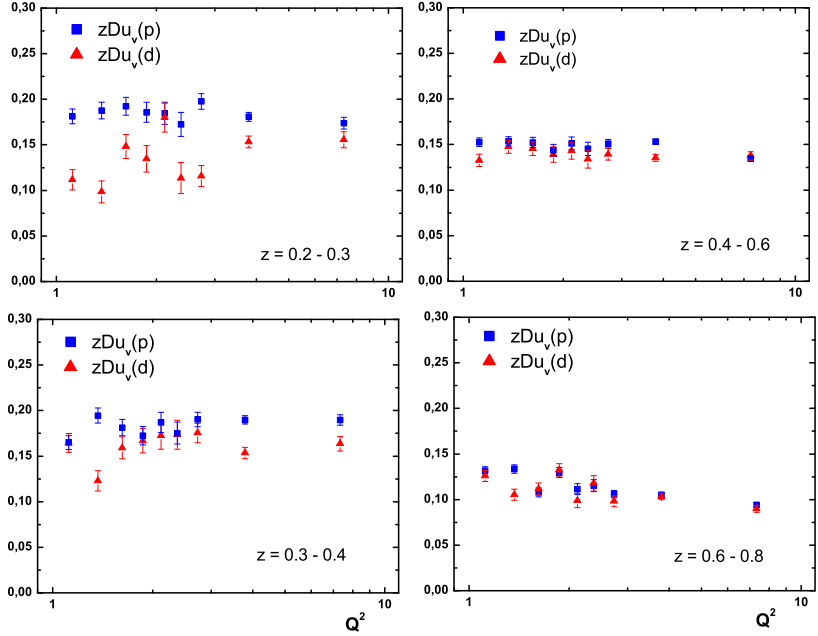}
\caption{\footnotesize Comparison between the LO QCD nonsinglet
fragmentation functions $zD^{\pi^{+}}_{u_v}(z, Q^2)$ extracted
from the differences of the pion multiplicities. Blue (red) points
correspond to the data using a proton (deuteron) target,
respectively. } \label{fig1}
\end{center}
\end{figure}

We have performed the same analysis in NLO QCD using for the
nonsinglet semi-inclusive structure functions Eqs.
(\ref{F1p}-\ref{FLd}) and the NLO QCD expressions for the usual
nucleon structure functions ($F_1,~F_L$). For the unpolarized PDFs
the NLO MRST'02 set \cite{MRST02} was used. In the NLO case one
can not extract the nonsinglet fragmentation function
$D^{\pi^{+}}_{u_v}(z, Q^2)$ directly from the data, and we have to
parametrize it at some fixed value of $Q^2$. The following
parametrization for $D^{\pi^{+}}_{u_v}(z, Q^2)$ at $Q^2_0=1~GeV^2$
was used:
\begin{equation}
zD^{\pi^{+}}_{u_v}(z, Q^2_0)=A_uz^{a_u}(1-z)^{b_u}[1+\gamma_u
(1-z)^{\delta_u}], \label{Duv_parQ0}
\end{equation}
where the parameters $\{A_u,a_u,b_u,\gamma_u,\delta_u \}$ are free
parameters to be determined from the fit to the data. The double
Mellin transform technique \cite{Double_Mellin} has been used to
calculate the nonsinglet semi-inclusive structure functions Eqs.
(\ref{F1p}-\ref{FLd}) from their moments.

We have found that using only the very tiny statistical errors one
can not achieve a satisfactory description of the data. We obtain
the following values for $\chi^2$ per point: $\chi^2(p)/Nrp=2.26$
from the fit to proton data, $\chi^2(d)/Nrp=1.96$ from the fit to
deuteron data, and $\chi^2(p)/Nrp=5.01$, $\chi^2(d)/Nrp=6.49$ from
the {\it combined} fit to the proton and deuteron data on the
differences of pion multiplicities
$M_{N}^{\pi^{+}-\pi^{-}}(x,Q^2,z),~ (N=p,d)$. We would like to
mention, however, that the statistical errors for the HERMES data
are between two and three times smaller than the systematic ones.
So, to obtain reasonable results from the fits to the data, the
systematic errors in this case have to be taken into account.

In Fig. 2 the best fit curves (solid lines) of our NLO QCD {\it
combined} fit to the proton and deuteron data on the differences
of pion multiplicities using the total errors are compared with
the data. The best fit curves (dashed lines) obtained from the
separate fits to the proton and deuteron data are also presented.
\begin{figure} [ht]
\begin{center}
  \includegraphics[height=.46\textheight]{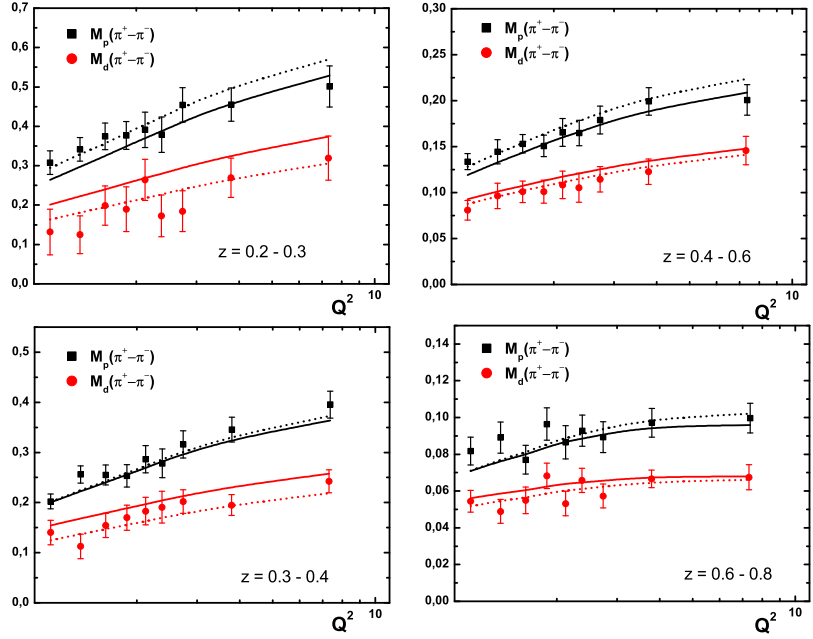}
\caption{\footnotesize Comparison between the data and the best
fit curves (black and red solid lines) obtained from a combined
NLO QCD fit to the proton and deutron data on the differences of
the pion multiplicities. The dashed curves correspond to the
separate fits to the proton and deuteron data. The errors of the
data are combined statistical and systematic.} \label{fig2}
\end{center}
\end{figure}
The numerical results for the combined fit are given in Table 1.
While in the case of separate fits to the proton and deuteron data
an excellent description for {\it all} z-bins of the proton as
well for the deuteron data is achieved:$\chi^2(p)/DOF=0.65$ and
$\chi^2(d)/DOF=0.50$ (see the dashed curves in Fig. 2), in the
combined fit to the data the situation is different. A good
description of the data is achieved except for the lower z bins,
$[0.2 <z < 0.3]$ and $[0.3 <z < 0.4]$, for the deuteron target,
for which $\chi^2/Nrp=2.04$ and $\chi^2/Nrp=1.24$, respectively
(see Table 1).
\begin{center}
\begin{tabular}{cl}
&{\bf Table 1.} Results of the NLO QCD combined fit to the proton
and deuteron \\& data on the differences of pion multiplicities.
The values of $\chi^2$ per point are\\
& presented for all z bins as well as separately for each z bin.
\end{tabular}
\vskip 0.6 cm
\begin{tabular}{|c|c|c|c|c|c|c|} \hline
 ~~Data~~~~~~&~~~$N_{data}$~~~~&~~~$\chi^2(proton)$~~~
 &~~~$\chi^2(deuteron)$~~~ \\ \hline
 All z-bins  &  36 &   0.79  &   1.19\\ \hline
 $z_1$-bin   &   9 &   0.96  &   2.04  \\
 $z_2$-bin   &   9 &   0.84  &   1.24   \\
 $z_3$-bin   &   9 &   0.59  &   0.59   \\
 $z_4$-bin   &   9 &   0.79  &   0.89   \\ \hline
\end{tabular}
\end{center}

One can see also in Fig. 2  that for these two bins the best fit
curves (solid red lines) lie systematically higher then the
central values of the data. Note that in the case of separate
fits, it was enough to use 3 free parameters for the input
parametrization of the nonsinglet structure function
$zD^{\pi^{+}}_{u_v}(z, Q^2_0)$ ($\gamma=0$ in Eq.
(\ref{Duv_parQ0})), while in the combined fit a better description
of the data was achieved using all 5 free parameters.
\begin{figure} [h]
\centerline{ \epsfxsize=3.2 in\epsfbox{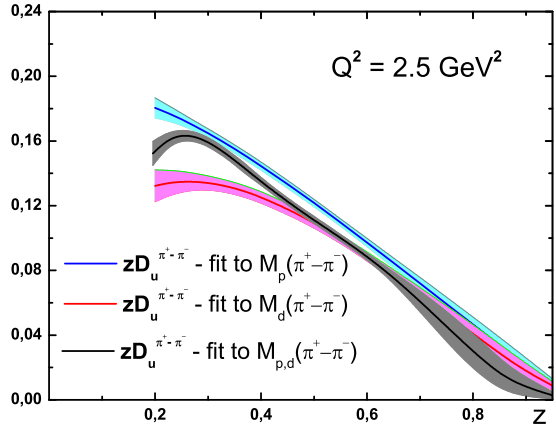} } \caption{
\footnotesize The nonsinglet fragmentation function
$zD_u^{(\pi^{+}-\pi^{-})}(z)$ at $Q^2=2.5~GeV^2$ extracted from
NLO QCD fits to a) proton, b) deuteron and c) proton and deuteron
data on the differences of pion multiplicities.} \label{fig3}
\end{figure}
The extracted nonsinglet fragmentation function
$zD_u^{\pi^{+}-\pi^{-}}(z,Q^2)$ from the combined NLO QCD fit to
the proton and deuteron data is presented in Fig. 3 as a function
of $z$ for $Q^2 = 2.5~GeV^2$ (black curve) together with its error
band, and in Fig. 4 as a function of $Q^2$ at any fixed z-bin for
the measured $Q^2$ values. In both the figures it is compared with
the nonsinglet FFs extracted from the separate fits to the proton
(blue curves) and deuteron (red curves) data. In Fig. 4 the LO
nonsinglet FFs extracted directly from the proton (blue points)
and deuteron (red points) data, are also presented.
\begin{figure} [h]
\begin{center}
  \includegraphics[height=.46\textheight]{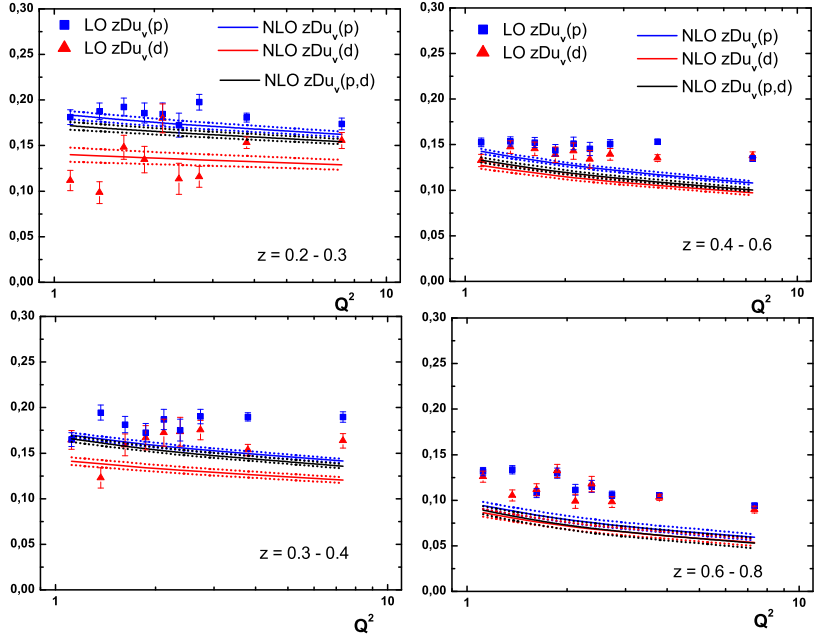}
\caption{\footnotesize Comparison between the LO and NLO
nonsinglet fragmentation functions $D^{\pi^{+}}_{u_v}(Q^2)$ at the
different z-bins. For details see the text.} \label{fig4}
\end{center}
\end{figure}

So, one can conclude from our NLO results that the combined fit to
the proton and deuteron data on the differences of the pion
multiplicities confirm the observation found in the LO analysis:
Assuming $SU(2)$ symmetry for the pion fragmentation functions and
applying the factorized QCD treatment to the HERMES $[Q^2,z]$ data
leads to problems with the description of the data for the lowest
z bins. In addition, an important fact coming from the NLO
analysis is that the problems are connected {\it only} with the
deuteron data (see Table 1 and red curves in Fig. 2). We have no
explanation of this point.

\subsection{SU(2) symmetry breaking}

One way to try to achieve a better fit to the data in the low z
bins too is to suppose that isospin SU(2) symmetry is broken. In
this case Eq. (\ref{SU2}) does not hold and there are two
independent nonsinglet fragmentation functions
$D_u^{\pi^{+}-\pi^{-}}(z,Q^2)$ and $D_d^{\pi^{+}-\pi^{-}}(z,Q^2)$
to be determined simultaneously from the NLO combined fit to the
proton and deuteron data, using for the semi-inclusive structure
functions Eqs. (3-6). The new nonsinglet
$D_d^{\pi^{+}-\pi^{-}}(z,Q^2) \equiv D_{d_v}^{\pi^{+}}(z,Q^2)$ is
parametrized at $Q^2_0=1~GeV^2$ like $D_{u_v}^{\pi^{+}}$ (see Eq.
(\ref{Duv_parQ0})) with free parameters
$\{A_d,a_d,b_d,\gamma_d,\delta_d \}$. The equality $a_d = a_u$ for
the parameters $a_u$ and $a_d$ has been used in the fit. The
results of the fit are illustrated in Table 2 and Fig.5.
\begin{center}
\begin{tabular}{cl}
&{\bf Table 2.} Results of the NLO QCD combined fit to the proton
and deuteron \\& data on the differences of pion multiplicities
when $SU(2)$ symmetry is broken. \\
&The values of $\chi^2$ per point are presented for all z bins as
well as separately for \\ & each z bin. The values in the brackets
correspond to $SU(2)$-symmetry fit.
\end{tabular}
\vskip 0.6 cm
\begin{tabular}{|c|c|c|c|c|c|c|} \hline
 ~~Data~~~~~~&~~~$N_{data}$~~~~&~~~$\chi^2(proton)$~~~
 &~~~$\chi^2(deuteron)$~~~ \\ \hline
 All z-bins  &  36 & 0.82 (0.79) & 0.50 (1.19)\\ \hline
 $z_1$-bin   &   9 & 0.86 (0.96) & 0.83 (2.04) \\
 $z_2$-bin   &   9 & 0.77 (0.84) & 0.44 (1.24) \\
 $z_3$-bin   &   9 & 0.81 (0.59) & 0.14 (0.59) \\
 $z_4$-bin   &   9 & 0.85 (0.85) & 0.60 (0.89) \\ \hline
\end{tabular}
\end{center}
\begin{figure} [h]
\begin{center}
  \includegraphics[height=.42\textheight]{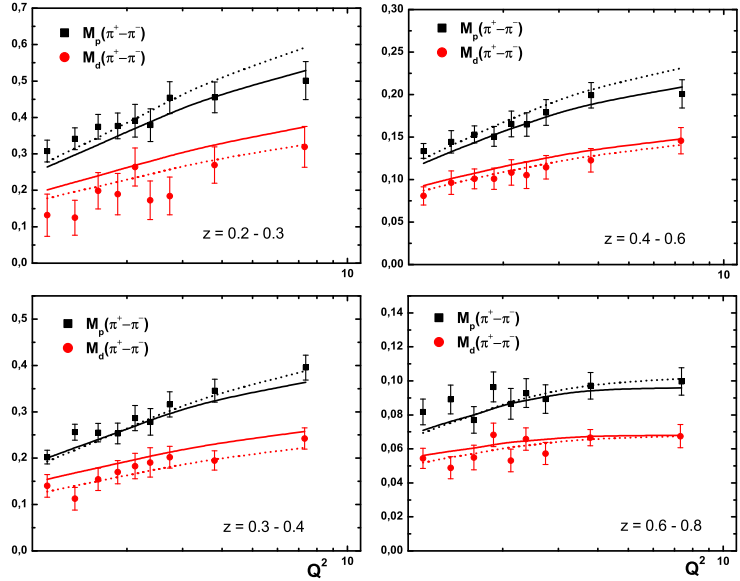}
\caption{\footnotesize Comparison between the data and the best
fit curves (black and red dashed lines) obtained from a combined
NLO QCD fit to the proton and deutron data on the differences of
pion multiplicities without $SU(2)$ symmetry. The solid curves
correspond to the $SU(2)$ symmetric fit to the same data.}
\label{fig5}
\end{center}
\end{figure}

As seen from Table 2 and Fig. 5, a significant improvement in the
fit to the deuteron data for the first two z-bins is achieved and
a good description of both the proton and deuteron data is
obtained. However, as seen from Fig. 6, where the extracted
nonsinglet FFs $D_u^{\pi^{+}-\pi^{-}}(z,Q^2)$ and
$D_d^{\pi^{+}-\pi^{-}}(z,Q^2)$ are presented, the violation of
$SU(2)$ symmetry is at a level, especially for the values of $z <
0.4$, which is hardly credible. The relation between the two
nonsinglets at $Q^2 = 1~ GeV^2$ obtained from the fit to the data
is:
\begin{equation}
D_d^{\pi^{+}-\pi^{-}}(z)=-2.18(1-z)^{0.46}D_u^{\pi^{+}-\pi^{-}}(z)
\end{equation}
Remember that when SU(2) symmetry holds:
$D_d^{\pi^{+}-\pi^{-}}(z)=-D_u^{\pi^{+}-\pi^{-}}(z)$.
\begin{figure}[ht]
\centerline{ \epsfxsize=3.2in\epsfbox{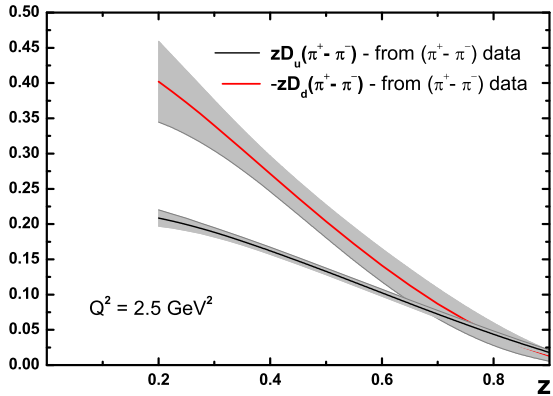} } \caption{
\footnotesize Nonsinglet fragmentation functions
$zD_u^{(\pi^{+}-\pi^{-})}(z)$ and $zD_d^{(\pi^{+}-\pi^{-})}(z)$ at
$Q^2=2.5~GeV^2$ together with their error bands extracted from a
NLO QCD combined fit to proton and deuteron data on the
differences of pion multiplicities.} \label{fig6}
\end{figure}

\section{Comments}

We think that what appears as a non-credible violation of isospin
symmetry could be a consequence of using the factorized QCD
treatment of the data in kinematic region where it is unjustified.
Indeed, according to Berger's phenomenological criterium \cite
{Berger} proposed a long time ago, there is a strong correlation
between $W$, the invariant mass of the hadrons $X$ produced in the
fully inclusive lepton nucleon process $l+N \rightarrow l+X$, and
the region z, where one can clearly separate the quark and target
fragmentation effects, i.e. where the hard scattering and the
hadronization factorize and the usual factorized QCD treatment is
valid. It was shown in \cite {Berger} that the  smaller the values
of z, the larger the values of W have to be for a clean separation
between the current and target jets. In \cite{Mulders} plots (Fig.
2 and Fig. 3) are presented showing the z-values for which it is
probably safe to use the factorized approach, for a quark
fragmenting into different hadrons $(\pi, K, N, \Lambda)$,
produced in  SIDIS processes, for  values of $W=5$ and $W=20$. As
seen from these plots, in the case of pions, values of $W\geq
5~GeV$ are needed for a clean separation of the quark and target
fragmentation effects in the data, and an unambiguous extraction
of the collinear fragmentation functions $D_{q,\bar
q}^{\pi^{+}}(z,Q^2)$ for $z \geq 0.2$. Keeping in mind that for
the HERMES data on pion multiplicities in the region $0.2 \leq z
\leq 0.8$ the corresponding values of $W$ are in the region $4.1 <
W < 4.5$, one sees that Berger's criterium is not satisfied for
the smaller values of z, and certainly for the first z-bin,
$0.2\leq z \leq 0.3$.

Finally, we would like to mention that in the $SU(2)$-symmetry
case we have found good agreement between the nonsinglet
$D_u^{\pi^{+}-\pi^{-}}(z,Q^2)$ extracted directly from the data on
the differences of the pion multiplicities
$M_{p(d)}^{\pi^{+}-\pi^{-}}(x,Q^2,z)$ and that obtained as a
difference between the favored $D_u^{\pi+}$ and unfavored
$D_u^{\pi-}=D_{\bar u}^{\pi+}$ fragmentation functions extracted
from our fit to $M_{p,d}^{\pi+}$ and $M_{p,d}^{\pi-}$ data
\cite{DSPIN'13}. The nonsinglets extracted in these two different
ways are shown in Fig. 7 (note that the highest value of $z$ for
the data is 0.7). The agreement shown in Fig (7) confirms that the
usual assumption about the unfavored fragmentation functions
\begin{equation}
D_s^{\pi+}=D_{\bar s}^{\pi+}=D_{\bar u}^{\pi+}
\label{unfavFFs}
\end{equation}
made in our fit \cite{DSPIN'13} to the $M_{p,d}^{\pi+}$ and
$M_{p,d}^{\pi-}$ data, is acceptable, but at the same time the
agreement is surprising given the problems reported earlier about
fitting the deuteron data on the difference of pion multiplicities
in the lowest z bins, when imposing isospin invariance.
\begin{figure}[ht]
\centerline{ \epsfxsize=3.2in\epsfbox{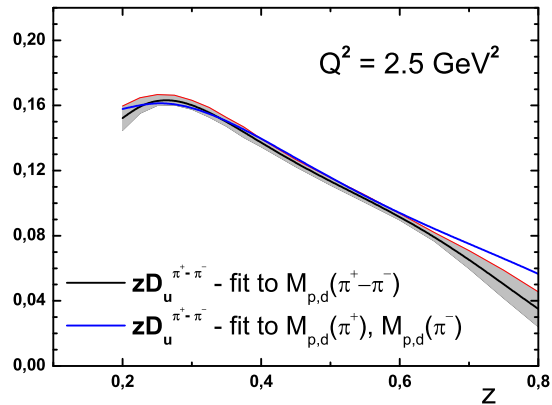} } \caption{
\footnotesize Comparison between nonsinglet fragmentation function
$zD_u^{(\pi^{+}-\pi^{-})}(z)$ extracted from the fit to proton and
deuteron data on the differences of the pion multiplicities  and
that one constructed from favored $zD_u^{\pi+}$ and unfavored
$zD_u^{\pi-}$ fragmentation functions determined from a NLO
combined fit to proton and deuteron data on pion multiplicities
themselves.} \label{fig7}
\end{figure}

Our NLO {\it favored} and {\it unfavored} pion FFs, extracted from
the HERMES data on multiplicities $M_{p,d}^{\pi+}$ and
$M_{p,d}^{\pi-}$ have been discussed in our paper \cite {DSPIN'13}
and compared to those determined by HKNS (Hirai, Kumano, Nagai,
Sudoh)
\cite {HKNS} and DSS (de Florian, Sassot, Stratmann) \cite {DSS}
which were obtained respectively from the semi-inclusive $e^+ \,
e^-$ annihilation data alone, and from the global fit to the
semi-inclusive $e^+ \, e^-$ annihilation data, the data on
single-inclusive hadron production in hadron-hadron collisions and
the {\it unpublished} HERMES'05 data on semi-inclusive
lepto-production of hadrons.

In Fig. 8 we compare our NLO nonsinglet fragmentation function
$zD_u^{(\pi^{+}-\pi^{-})}(z)$ extracted \emph{directly} from the
HERMES data on the difference between $\pi^{+}$ and $\pi^{-}$
multiplicities to those of HKNS and DSS obtained as differences
between their favored $D_u^{\pi+}$ and unfavored $D_{\bar
u}^{\pi+}=D_u^{\pi-}$ NLO fragmentation functions. Note that in the
DSS analysis $SU(2)$- symmetry was broken and Eq. (\ref{SU2}) does
not hold. That is why the nonsinglet fragmentation function
$-zD_d^{(\pi^{+}-\pi^{-})}(z)$ for DSS is also presented in Fig. 8.
As seen from Fig. 8, the discrepancy between the nonsinglets, is
significant.
\begin{figure}[h]
\centerline{ \epsfxsize=3.2in\epsfbox{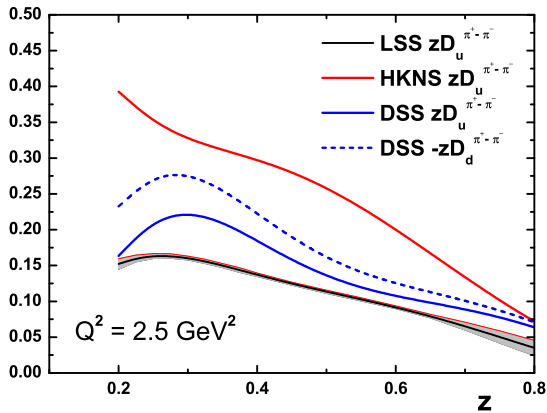} } \caption{
\footnotesize Comparison between our NLO nonsinglet fragmentation function
$zD_u^{(\pi^{+}-\pi^{-})}(z)$ extracted from the fit to proton and
deuteron data on the differences of the pion multiplicities  and
those of HKNS and DSS constructed from their favored $zD_u^{\pi+}$
and unfavored $zD_u^{\pi-}$ NLO fragmentation functions. \label{fig8}}
\end{figure}

It is important to mention that the semi-inclusive 
$e^+ \, e^-$ annihilation data give no information how to disentangle 
qiark $D_q^h(z, Q^2)$ from anti-quark $D_{\bar q}^h(z, Q^2)$ 
fragmentation, and only their sum $D_q^h + D_{\bar q}^h$ can be 
determined from the data, while the important role of the 
semi-inclusive DIS processes is that they allow to separate 
$D_q^h(z,Q^2)$ from $D_{\bar q}^h(z, Q^2)$. We think that this is
the reason for the large difference between our nonsinglet FF and
the HKNS one.  As for the inconsistency between our and the DSS
nonsinglet FFs, that is a consequence of the fact that the DSS
group has used in their analysis the unpublished HERMES'05 data
which are not consistent with the final HERMES data which we have
used \cite {HERMES}.

We feel it is important to mention a totally different issue which
might be the source of the entire problem in the perturbative QCD
description of the HERMES data, namely  the fact that there
appears to be an inconsistency between the two ways HERMES has
chosen to present their data. It appears to us that the $[x,z]$
and the $[Q^2,z]$ presentations are incompatible \cite{DSPIN'13}.
This could be a signal that maybe there is something wrong with
the data.

\begin{center}
{\bf Acknowledgments}
\end{center}

We are grateful to A. Kotzinian for the useful comments and
helpful discussions. This research was supported by the
JINR-Bulgaria Collaborative Grant, and by the RFBR Grants (Nrs
12-02-00613, 13-02-01005 and 14-01-00647).


\end{document}